Method of model reduction and multifidelity models for solute transport

in random layered porous media

Zhijie Xu[a] and Alexandre M. Tartakovsky

Computational Mathematics Group

Physical & Computational Sciences Directorate

Pacific Northwest National Laboratory

[a] Electronic mail: zhijie.xu@pnnl.gov Tel: 509-372-4885

**Abstract**

This work presents a hierarchical model for solute transport in bounded layered porous media with random permeability. The model generalizes the Taylor-Aris dispersion theory to stochastic transport in random layered porous media with a known velocity covariance function. In the hierarchical model, we represent (random) concentration in terms of its cross-sectional average and a variation function. We derive a one-dimensional stochastic advection-dispersion-type equation for the average concentration and a stochastic Poisson equation for the variation function, as well as expressions for the effective velocity and dispersion coefficient. We observe that velocity fluctuations enhance dispersion in a non-monotonic fashion: the dispersion initially increases with correlation length $\lambda$, reaches a maximum, and decreases to zero at infinity. Maximum enhancement can be obtained at the correlation length about 0.25 the size of the porous media perpendicular to flow.

**I. Introduction**

Many scientific applications (e.g., barotropic flow, contaminant transport, and functionally graded materials) are multiscale and stochastic in nature with uncertainties stemming from random initial and/or boundary conditions and/or stochastic parameter fields. Solving these stochastic problems is both theoretically and computationally challenging.

Existing perturbation-based moment methods for solving stochastic advection-dispersion equations develop non-physical bi-modal behavior for average concentration [1, 2]. The moment solution based on the macro-dispersion theory [3] requires knowledge of Green's function, which is expensive to compute numerically and can only be found analytically for a small class of problems (e.g., infinite domains). Another drawback to these methods is their accuracy rapidly deteriorates with increasing variance of the random parameters (i.e., advection velocity and/or dispersion coefficient). Other statistical approaches, including Monte Carlo (MC) methods, suffer from a low convergence rate ($O(N^{-1/2})$, where N is the number of samples) and are destined to fail when directly applied to problems with large numbers of degrees of freedom [4].

Polynomial chaos (PC)-based methods [5-7] currently are a method of choice for quantifying uncertainty [8-10]. However, these methods suffer from the so-called "curse of dimensionality" and become prohibitively expensive when applied to problems with correlated-in-space random parameters characterized by small correlation length and/or large variance [11-17].

In this paper, we present a new reduction method for solute transport in layered porous media with random distribution of the hydraulic conductivity across the layers.

We derive stochastic equations for the spatial average of concentration and variations around the average. The spatial average represents the large-scale concentration and is governed by a stochastic advection-dispersion equation with the effective stochastic advection velocity and dispersion coefficient. The small-scale variability of the concentration, caused by the small-scale velocity fluctuations, is captured by the variation function, depending on the velocity covariance. The resulting hierarchical stochastic models enable efficient solution of the original problem with significantly reduced dimensionality.

Aris and Taylor's classical dispersion theory was developed for long-time evolution of solute concentration (Taylor [18, 19] and Aris [20], see also Philip [21], Brenner [22], Gill and Sankaras [23], Smith [24], Frankel and Brenner [25], Fischer [26], and Xu [27, 28]). Whitaker, Adler and Brenner, and Bear later generalized this theory to (deterministic) flow in porous media. Neuman [3] and Koch and Brady [29] derived deterministic effective dispersion equations for solute transport in the stochastic velocity field. Our method generalizes Taylor dispersion theory [18, 19] for transport in the stochastic velocity field. Unlike Neuman's macro-dispersion theory [3] (which results in deterministic "macroscale" equations), our method yields a stochastic "macroscale" advection-dispersion equation and an expression for microscale concentration fluctuations. A stochastic form of the effective equation allows efficient uncertainty quantification and parameter and state estimation using small-scale concentration measurements.

In the proposed method, for known statistics of the advective velocity, the variation function, effective advection velocity, and effective dispersion coefficient can be computed analytically. The stochastic parameters in the effective equation have smaller variance and larger correlation lengths than their small-scale counterparts in the original advection-dispersion equations. Therefore, the effective stochastic equation can be solved using Monte Carlo simulations (MCS) with significantly coarser resolution than the one required in the MCS solution for the original equation. In addition, as the accuracy of these methods increases with decreasing variances of the random parameters in the stochastic equations, the moment equation and macro-dispersion methods should be more accurate for the effective equation than for the original equations.

## II. Formulation of the Model

Here, we consider solute transport in porous media consisting of homogeneous layers with random permeability distribution across the layers. The randomness in permeability leads to randomness of the advective velocity. The two-dimensional (2-D) geometry of the problem is defined in Figure 1. The flow domain is bounded in the $y$ direction ($a$ is the size of the domain in the $y$ direction) and is infinite in the $x$ direction. Conservative solute transport in this domain can be described by the 2-D advection-dispersion equation

$$\partial c/\partial t + v \cdot \nabla c = D \nabla^2 c \tag{1}$$

subject to no flux boundary conditions

$$\left.\frac{\partial c}{\partial y}\right|_{y=0,a} = 0 \tag{2}$$

at the top and bottom of the domain. The advection velocity satisfies Darcy's law, $v(y) = -K(y)\partial h/\partial x$, where $K(y)$ is random conductivity and $\partial h/\partial x$ is the constant (in time and space) head gradient. In the preceding equations, $c(x, y, t)$ is the solute concentration at position (*x, y*) and time *t*, and *D* is the dispersion coefficient assumed here to be constant and the same for each layer. For transport in a channel (*v(y)* having a parabolic profile), Taylor derived an analytical expression for the dispersion coefficient [18]. Here, we derive an expression for the dispersion coefficient for random velocity *v(y)* with the prescribed mean, variance, and correlation function. For an infinite domain in the *x* direction or a domain with the length *L*, such as $L \gg a$, we may write the total concentration as [18, 30, 31]

$$c(x,y,t) \approx \bar{c}(x,t) + \eta(y)\frac{\partial \bar{c}}{\partial x}, \tag{3}$$

where $\bar{c}(x,t)$ is the cross-sectional average of total concentration $c(x,y,t)$ and the cross-sectional averaging operator $\overline{\square}$ is defined as:

$$\overline{\square} = \frac{1}{a}\int_0^a (\square) dy. \tag{4}$$

The function $\eta(y)$ is a measure of the velocity variation along the *y* direction and will be derived later. Equation (3) decomposes the total stochastic concentration solution $c(x,y,t)$ in terms of the cross-sectional average concentration $\bar{c}$ and its first-order gradient $\partial \bar{c}/\partial x$. Though higher-order expression for the correction (in terms of the spatial derivatives of $\bar{c}(x,t)$) may be obtained [27, 32], only the first-order correction is considered in this study. The total uncertainty in solution $c(x,y,t)$ can be further decomposed into the ensemble contribution in average solution $\bar{c}(x,t)$ and configurational contribution in variation function $\eta(y)$ [30].

Similarly, the total velocity field is decomposed into the cross-sectional average $\bar{v}$

$$\bar{v} = \frac{1}{a}\int_0^a v(y)dy \tag{5}$$

and velocity fluctuation $v'$ around that average

$$v(y) = \bar{v} + v'(y), \tag{6}$$

where both $\bar{v}$ and $v'$ are random. The velocity fluctuation $v'$ has a zero cross-sectional average:

$$\overline{v'} = \frac{1}{a}\int_0^a v'(y)dy = 0. \tag{7}$$

The zero ensemble average $\langle v' \rangle = 0$ is satisfied only if the ensemble average $\langle v(y) \rangle = \langle v \rangle$ is independent of $y$, which is assumed in the current study.

The key part of the proposed solution method for stochastic partial differential equation (PDE) (1) is to formulate the equations and solutions for cross-sectional average concentration $\bar{c}(x,t)$ and in-plane variation function $\eta(y)$. The boundary condition of $c(x,y,t)$ (Eq. (2)) immediately leads to the boundary conditions for $\eta(y)$:

$$\left.\frac{\partial \eta}{\partial y}\right|_{y=0,a} = 0. \tag{8}$$

By applying the cross-sectional operator to both sides of Eq. (3), the cross-sectional average of function $\eta(y)$ is found as:

$$\bar{\eta} = \frac{1}{a}\int_0^a \eta(y)dy = 0. \tag{9}$$

Substitution of Eq. (3) into the original stochastic PDE (1) and applying the cross-sectional average operator to both sides of Eq. (1) lead to the equation for $\bar{c}(x,t)$:

$$\frac{\partial \overline{c}}{\partial t} + \left( \overline{v} - D\overline{\frac{\partial^2 \eta}{\partial y^2}} \right) \cdot \frac{\partial \overline{c}}{\partial x} = \left( D - \overline{v\eta} \right) \frac{\partial^2 \overline{c}}{\partial x^2}. \qquad (10)$$

It is evident that the reduced model for $\overline{c}(x,t)$ (Eq. (10)), a one-dimensional stochastic PDE, is easier to solve than the original Eq. (1). According to Eq. (7), the statistical ensemble average of velocity fluctuation is

$$\langle v'(y) \rangle = \frac{1}{a} \int_0^a \langle v(y) \rangle dy - \langle v(y) \rangle, \qquad (11)$$

where the operator $\langle \Box \rangle$ represents the statistical ensemble average of a field variable "g."

Using the boundary condition (8), we find the conditions for $\eta(y)$:

$$\overline{\frac{\partial^2 \eta}{\partial y^2}} = \frac{\partial \eta}{\partial y}\bigg|_{y=a} - \frac{\partial \eta}{\partial y}\bigg|_{y=0} = 0. \qquad (12)$$

By substituting Eq. (12) into Eq. (10), the equation for $\overline{c}(x,t)$ is further reduced to

$$\frac{\partial \overline{c}}{\partial t} + \overline{v} \cdot \frac{\partial \overline{c}}{\partial x} = \widetilde{D} \frac{\partial^2 \overline{c}}{\partial x^2}, \qquad (13)$$

where

$$\widetilde{D} = \left( D - \overline{v\eta} \right) = \left( D - \overline{v'\eta} \right) \qquad (14)$$

is a stochastic scalar function representing the effective dispersion coefficient for $\overline{c}(x,t)$ due to the random velocity $v(y)$. The term $\overline{v'\eta}$ represents the contribution of the non-uniform advection velocity field $v(y)$ to the dispersion.

Thus far, we have formulated the stochastic advection-dispersion equation (13) for $\overline{c}(x,t)$ with stochastic advection velocity $\overline{v}$ and stochastic effective dispersion $\widetilde{D}$ that depends on the in-plane function $\eta(y)$. To derive the equation for $\eta(y)$, Eqs. (3) and (13) are substituted into the original stochastic PDE (1), which leads to:

$$\frac{\partial \bar{c}}{\partial x}\left(v - \bar{v} - D\frac{\partial^2 \eta}{\partial y^2}\right) + \frac{\partial^2 \bar{c}}{\partial x^2}\left(v\eta - \bar{v}\eta - \overline{v\eta}\right) - \frac{\partial^3 \bar{c}}{\partial x^2}\overline{\eta v \eta} = 0 \qquad (15)$$

Because the expansion of total concentration $c(x,y,t)$ (Eq. (3)) only retains a first-order correction, we obtain an equation for $\eta(y)$ satisfying Eq. (15) to the first order:

$$D\frac{\partial^2 \eta}{\partial y^2} = v - \bar{v} = v'. \qquad (16)$$

By integrating Eq. (16) twice and using the boundary conditions (8) and constraint (9), we obtain the solution for $\eta(y)$:

$$\eta(y) = \frac{1}{D}\left[\int_0^y \int_0^{y_2} v'(y_1) dy_1 dy_2 - \frac{1}{a}\int_0^a \int_0^y \int_0^{y_2} v'(y_1) dy_1 dy_2 dy\right]. \qquad (17)$$

Taking ensemble average of both sides of Eq. (17), we find the necessary conditions for

$$\langle \eta(y) \rangle = 0 \qquad (18)$$

is $\langle v' \rangle = 0$.

The stochastic effective dispersion coefficient $\tilde{D}$ can be derived from Eq. (14). First, we integrate Eq. (17) by parts and the boundary condition for $\eta(y)$ in Eq. (8) to obtain

$$-\overline{v'\eta} = -D\overline{\frac{\partial^2 \eta}{\partial y^2}\eta} = D\overline{(\partial \eta/\partial y)^2}. \qquad (19)$$

Substituting this into Eq. (14), we obtain the solution for $\tilde{D}$:

$$\tilde{D} = D\left(1 + \overline{(\partial \eta/\partial y)^2}\right) = D\left(1 + \frac{1}{D^2}\overline{\int_0^y v'(y_1) dy_1 \int_0^y v'(y_2) dy_2}\right). \qquad (20)$$

It can be seen from Eq. (20) that the stochastic effective dispersion $\widetilde{D} \geq D$, i.e., the heterogeneity (fluctuations) in advective velocity $v(y)$ always enhances the effective dispersion.

Next, we demonstrate the consistency of our formulation with the Taylor-Aris theory for the (deterministic) parabolic velocity profile for $v(y)$:

$$v(y) = \frac{3}{2}\bar{v}\left(1 - \frac{y^2}{(a/2)^2}\right). \tag{21}$$

Substitution of Expression (21) into Eq. (17) leads to the corresponding solution for $\eta(y)$:

$$\eta(y) = \frac{\bar{v}a^2}{60D}\left[1 - \frac{15}{8}\left(1 - \left(\frac{y}{a/2}\right)^2\right)^2\right]. \tag{22}$$

Then, the effective dispersion coefficient $\widetilde{D}$ can be computed via substitution of Eqs. (21) and (22) into Eq. (14) as

$$\widetilde{D} = \left(D - \overline{v'\eta}\right) = D\left(1 + \frac{P_e^2}{210}\right), \tag{23}$$

where the Péclet number is defined as $Pe = a\bar{v}/D$. This result exactly recovers the Taylor's dispersion coefficient [18].

Next, we study statistical properties of $\bar{v}$ and $\widetilde{D}$ in Eq. (13). Mean and variance of $\bar{v}$ can be analytically obtained for a given covariance function of stochastic velocity $v(y)$. Here, we assume that $v(y)$ is statistically homogeneous and has the constant (ensemble) mean $\langle v \rangle$ and exponential covariance function,

$$\langle v(y_1)v(y_2)\rangle = \langle v(y)\rangle^2 + \sigma^2 \exp\left(-\frac{|y_1 - y_2|}{\lambda}\right), \tag{24}$$

where $\sigma^2$ is the variance of velocity fluctuation and $\lambda$ is the correlation length. Then, the ensemble mean and variance of $\bar{v}$ are given by [33]

$$\langle \bar{v}\rangle = \langle v\rangle \tag{25}$$

and

$$\sigma_{\bar{v}}^2 = \langle \bar{v}^2\rangle - \langle \bar{v}\rangle^2 = 2\sigma^2\mu^2\left(e^{-1/\mu} + 1/\mu - 1\right), \tag{25}$$

where $\mu = \lambda/a$ is the dimensionless correlation length. Figure 1 shows the variation of non-dimensional ratio $\beta = \sigma_{\bar{v}}^2/\sigma^2$ with the correlation length $\mu$, where $\beta$ approaches 1 with increasing correlation length $\mu$ or $\sigma_{\bar{v}}^2 \to \sigma^2$ when $\mu \to \infty$.

The statistical mean of the effective dispersion $\tilde{D}$ can be obtained from Eq. (20) as

$$\frac{\langle \tilde{D}\rangle}{D} = 1 + \frac{\gamma a^2 \sigma^2}{D^2}, \tag{26}$$

where

$$\gamma = \overline{\int_0^y \int_0^y \langle v'(y_1)v'(y_2)\rangle dy_1 dy_2}\bigg/(a\sigma)^2 \tag{27}$$

is a dimensionless number representing the effect of velocity fluctuation on mixing enhancement. The covariance function of velocity fluctuation $v'(y_1)$ can be related to the covariance function of $v(y)$ using Eq. (6) as:

$$\langle v'(y_1)v'(y_2)\rangle = \langle v(y_1)v(y_2)\rangle + \langle \bar{v}^2\rangle - \langle \bar{v}\cdot v(y_2)\rangle - \langle \bar{v}\cdot v(y_1)\rangle. \tag{28}$$

The final expression for $\gamma$ is obtained using Eq. (28) as

$$\gamma = 4\mu^4\left(e^{-1/\mu}-1\right)+4\mu^3-\frac{1}{3}\mu^2\left(e^{-1/\mu}+5\right)+\frac{1}{3}\mu \tag{29}$$

and plotted in Figures 2 and 3 as a function of µ. These figures show that $\gamma$ increases from zero to its maximum value $\gamma = 0.026$, corresponding to $\mu \approx 0.25$, then decreases to zero for large µ. Note that for the parabolic velocity profile (Eq. (21)), the velocity variance is

$$\sigma^2 = \overline{\left(v(y)-\overline{v}\right)^2} = \overline{v}^2/5, \tag{30}$$

and the equivalent $\gamma$ from Taylor's theory is $\gamma = 1/42 \approx 0.0238$ —only slightly smaller than the maximum value $\gamma=0.026$ for the random velocity. Maximum enhancement in mixing can be achieved for the stochastic velocity field $v(y)$ with a correlation length $\lambda \approx 0.25a$, where $a$ is the total layer thickness.

Finally, we perform MCS to compute the probability distribution functions (PDFs) of $\overline{v}$, $\overline{D}$, and $\eta(y)$. We assume the porous medium is made of 100 layers, and the velocity in each layer is constant with a uniform distribution defined on interval [0, 1]. Figures 4 and 5 illustrate the PDFs of $\overline{v}$ and $\overline{D}$ with $\overline{v}$ and $\widetilde{D}$ approaches Gaussian and $\chi^2$ distributions, respectively, for small correlation length µ.

Figure 6 depicts the realizations of $\eta(y)$ obtained from the MCS. The PDFs for $\eta(y)$ at the top ($y=1$) and middle ($y=0.5$) of the domain are presented in Figures 7 and 8. The PDF function for $\eta(y)$ approaches Gaussian distribution at all locations but with fluctuating variance that is larger at both upper and lower boundaries and smaller in the middle of the domain.

## II. Conclusions

We have presented a model reduction method that results in hierarchical stochastic models for solute transport in layered porous media with random distributions of advection velocity across different layers. The model, given by Eq. (3), approximates the concentration field $c(x,y,t)$ in terms of its cross-sectional average $\bar{c}(x,t)$ and in-plane variation function $\eta(y)$ (given by Eq. (16)), where $\bar{c}(x,t)$ represents the large-scale variability of $c(x,y,t)$ and is governed by the stochastic advection-dispersion equation (13) with effective advection velocity $\bar{v}$ (given by Eq. (5)) and effective dispersion coefficient $\widetilde{D}$ (given by Eqs. (14) or (20)). The small-scale variability in $c(x,y,t)$, caused by small-scale variability of the advection velocity $v(y)$, is captured by the in-plane function $\eta(y)$. The resulting hierarchical models can significantly reduce the problem of dimensionality for efficiently solving the original expensive problem. The effect of correlation field length $v(y)$ on the enhancement in dispersion also has been analytically examined. The maximum enhancement (maximum effective dispersion coefficient) was found for a correlation length at about $0.25a$. There is no enhancement (i.e., the effective dispersion coefficient is equal to the molecular diffusion coefficient) for both zero and infinity large correlation lengths.

**ACKNOLEDGEMENT**

This research was supported by LDRD program "Exploring Multilevel Numerical Methods for Extreme-scale Computing" from Pacific Northwest National Laboratory.

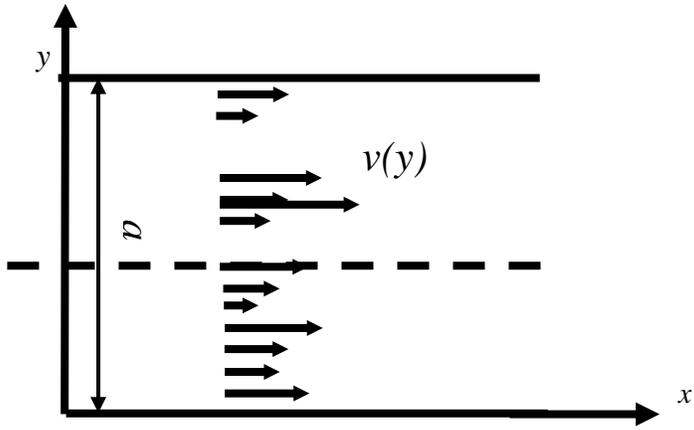
Figure 1. Flow confined by two parallel plates with a stochastic velocity profile *v(y)*.

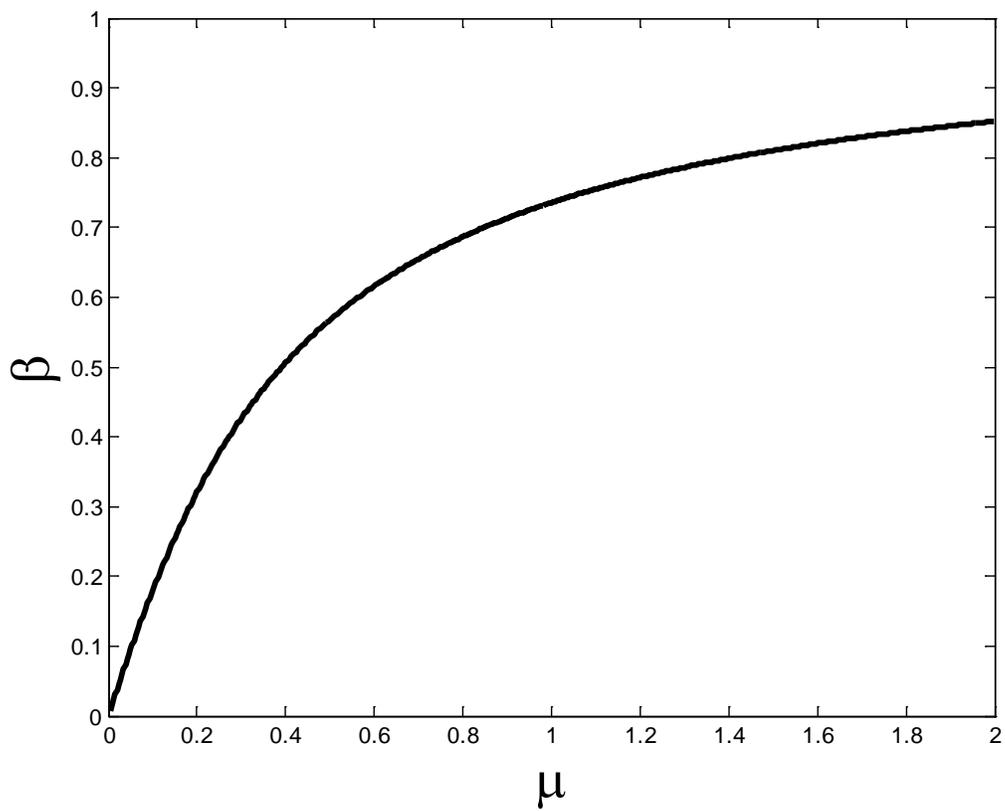

Figure 2. Fluctuation of variance ratio β with correlation length μ. The variance $\sigma_{\bar{v}}^2 < \sigma^2$ but approaches $\sigma^2$ when $\mu \to \infty$.

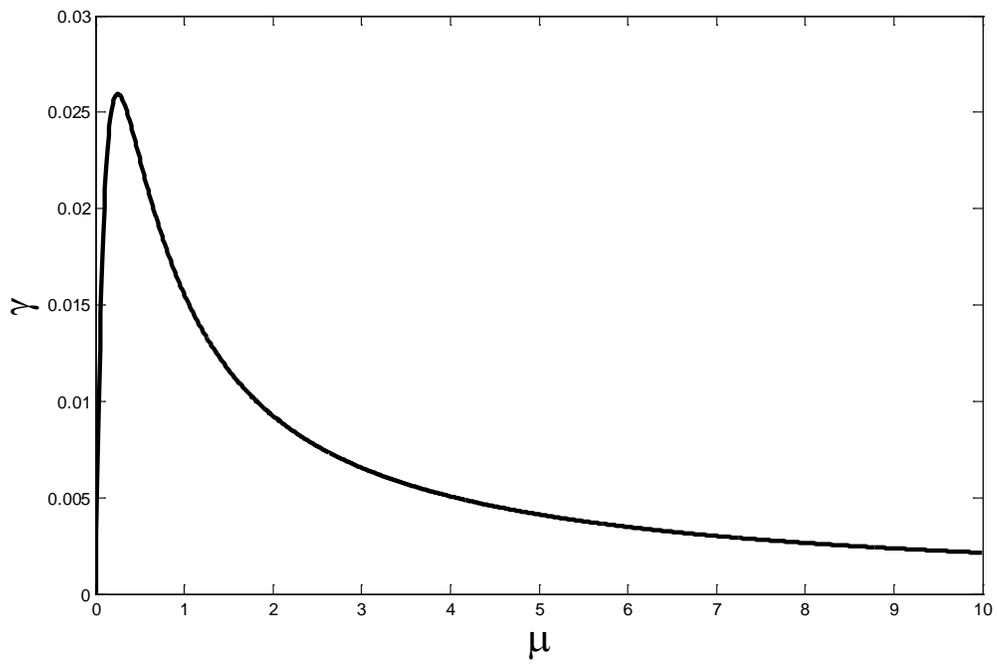

Figure 3. Variation of enhancement in dispersion with the correlation length.

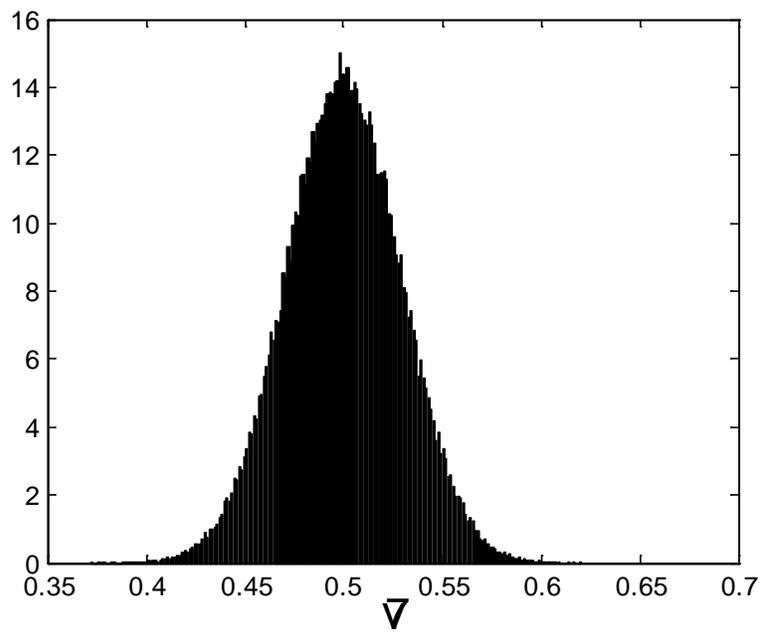

Figure 4. Probability density function of effective velocity $\bar{v}$.

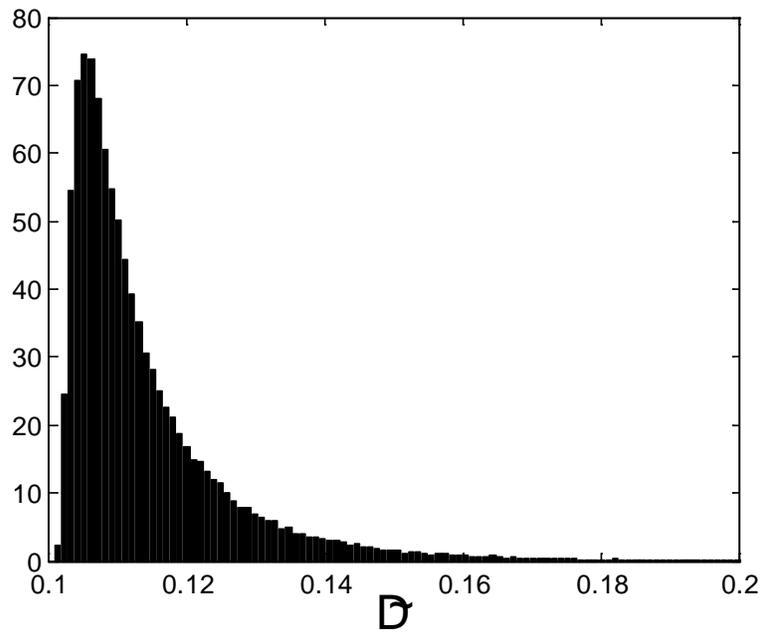

Figure 5. Probability density function of effective dispersion $\widetilde{D}$ with $D = 0.1$ corresponding to the dispersion of constant velocity. The ensemble mean $\langle \widetilde{D} \rangle$ shows the enhancement in dispersion due to velocity fluctuation.

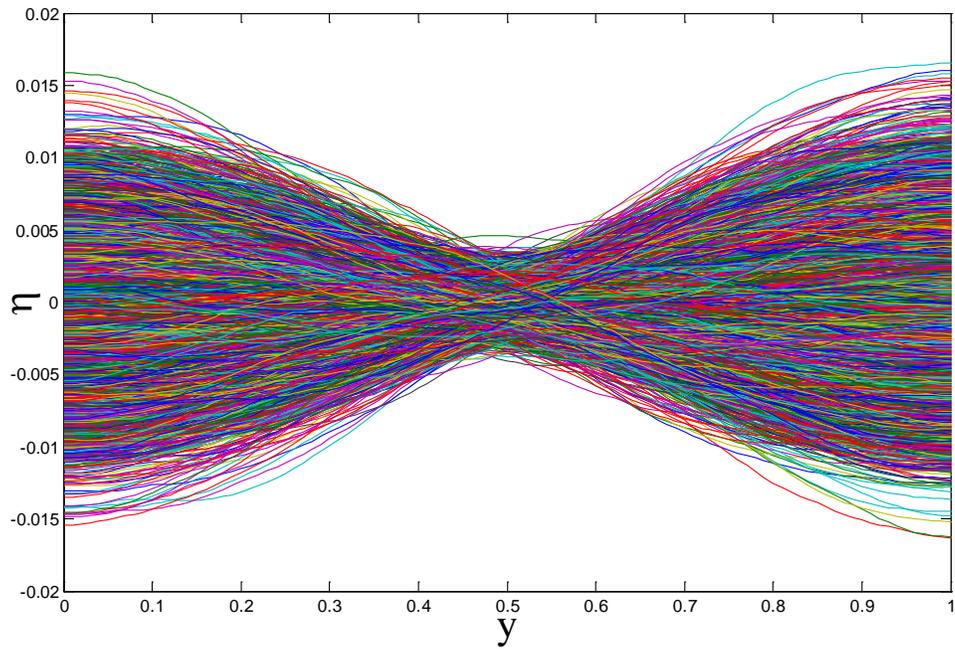

Figure 6. Plot of in-plane variation $\eta(y)$, fluctuating with $y$ from $10^5$ samplings.

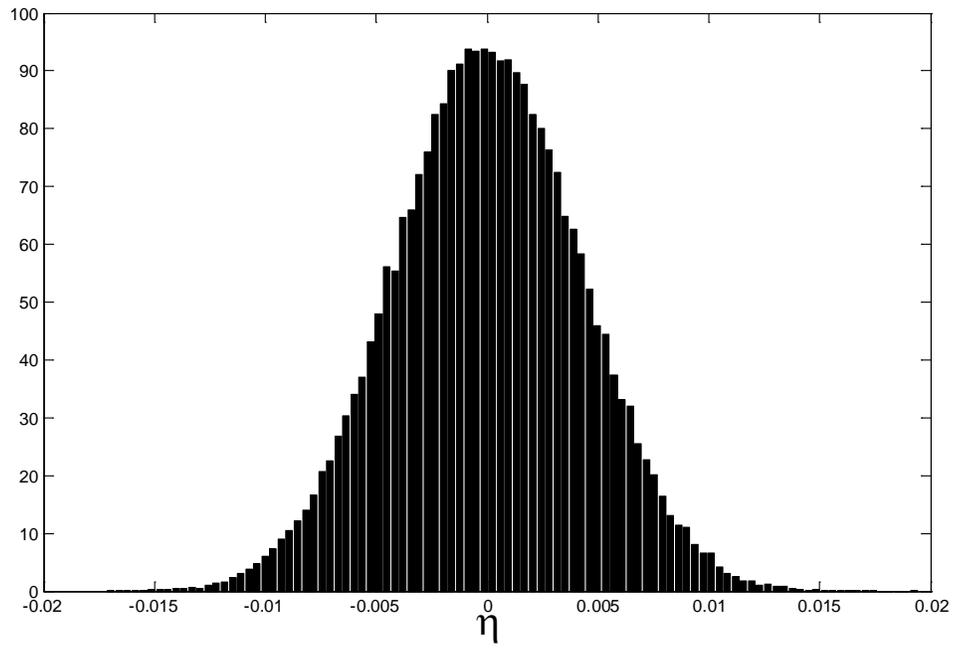

Figure 7. Probability density distribution of $\eta(y)$ at $y=1$.

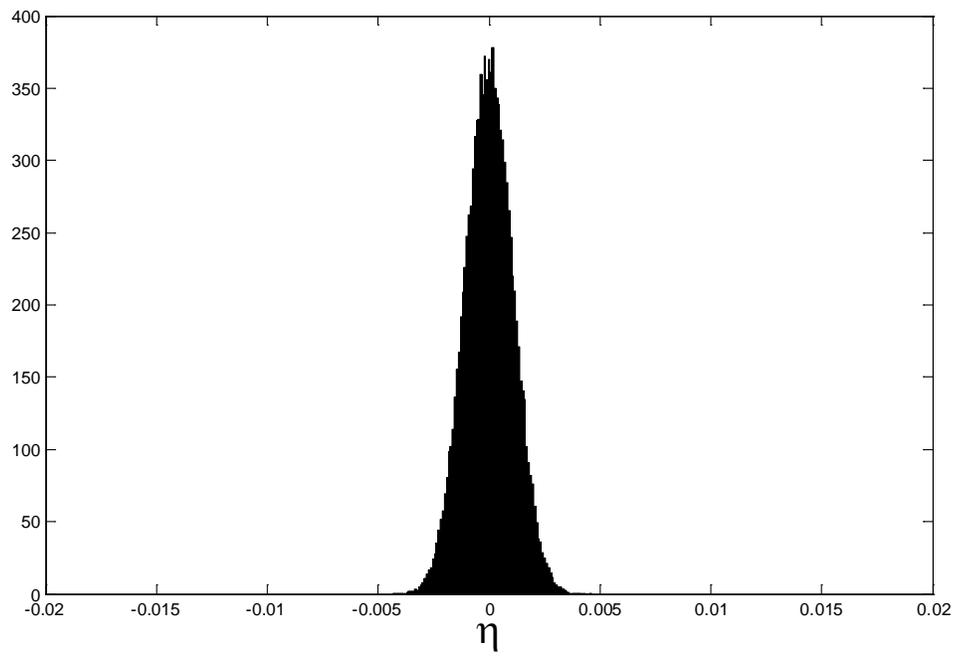

Figure 8. Probability density distribution of $\eta(y)$ at $y = 0.5$.